\documentclass{birkjour}

\usepackage[ansinew]{inputenc}
\usepackage{amsfonts}
\usepackage{lmodern}
\usepackage{mathrsfs}
\usepackage[all]{xy}
\usepackage{amsthm, amssymb, amsmath, latexsym}
\usepackage{color}
\usepackage{array, float}
\usepackage{bm}
\usepackage{hyperref}
\usepackage{graphicx}
 \graphicspath{{Images/}}
\newtheorem{Theorem}{Theorem}[section]

\theoremstyle{definition}

\theoremstyle{remark}

\numberwithin{equation}{section}

\begin{document}

%-------------------------------------------------------------------------------------------------------------------------------------------------------
% TITLE SECTION
%-------------------------------------------------------------------------------------------------------------------------------------------------------

\title{Lorentzian surfaces and the curvature of the Schmidt metric}

\author{Yafet Sanchez Sanchez}
\address{Mathematical Sciences and STAG Research Centre,\br University of Southampton,\br Southampton, SO17 1BJ.}
\email{E-mail:Y.SanchezSanchez@soton.ac.uk}

\author{Cesar Merlin}
\address{Centro Brasileiro de Pesquisas F\'isicas (CBPF),\br Rio de Janeiro,\br CEP 22290-180,\br Brazil.}
\email{E-mail:cmerlin@cbpf.br}

\author{Ricardo Reynoso Fuentes}
\address{Departamento de F\'isica, Facultad de Ciencias,\br Universidad Nacional Aut\'onoma de M\'exico (UNAM),\br Ciudad de M\'exico, M\'exico.}
\email{E-mail:rreynoso.92@ciencias.unam.mx}

\keywords{gravitational singularities, Lorentzian surfaces, Schmidt metric, general relativity, b-boundary }

\date{\today}

%-------------------------------------------------------------------------------------------------------------------------------------------------------
% ABSTRACT SECTION
%-------------------------------------------------------------------------------------------------------------------------------------------------------

\begin{abstract}

The $b$-boundary is a mathematical tool used to attach a topological boundary to incomplete Lorentzian manifolds using a Riemaniann metric called the Schmidt metric on the frame bundle. 
In this paper we give the general 
form of the Schmidt metric in the case of Lorentzian surfaces. Furthermore, we write the Ricci scalar of the Schmidt metric in terms of the Ricci scalar of the Lorentzian manifold and give some examples. Finally, we discuss some applications to general relativity. 
\end{abstract}

\maketitle
{\bf{PACS. }}{0420DW; 0240HW; 0420CV}
%-------------------------------------------------------------------------------------------------------------------------------------------------------
% INTRO SECTION
%-------------------------------------------------------------------------------------------------------------------------------------------------------

\section{Introduction}
One of the biggest surprises that General Relativity (GR) has given us is that under certain circumstances the theory predicts its
own limitations. There are two physical situations where we expect the theory to break down. The first one is the gravitational collapse 
of certain massive stars when their nuclear fuel is spent. The second one is the far past of the universe when the density and temperature 
were extreme. In both cases, we expect that the geometry of spacetime will show some pathological behaviour.

The nature of a gravitational singularity is a delicate issue. It might be tempting to define a gravitational singularity following 
other physical theories (such as electromagnetism) as the location where the relevant physical quantities are undefined. However, in 
the gravitational case this prescription does not work due to the identification of the spacetime background with the gravitational field. 
As a result, the concepts of `spatial location' and `temporal duration' have meaning only in the domain where the gravitational field is defined. 
This represents a problem because the size, place and shape of singularities can not be straightforwardly characterised by any physical measurement.

The first mathematical description of a gravitational singularity comes from Penrose and Hawking seminal theorems. They characterised singularities as obstructions to geodesic ompleteness and managed to 
show that this happen under certain conditions \cite{hawking1}.
Broadly speaking, the theorems establish that a spacetime ($\mathcal{M},g$) that satisfies simultaneously 
\begin{itemize}
  \item a condition on the curvature,
  \item an appropriate initial or boundary condition,
  \item and a global causal condition,
\end{itemize}
must be geodesically incomplete \cite{senovilla}.

One would like to attach a boundary to the incomplete spacetime to understand the singularity better. The procedure to attach a boundary to a Lorentzian manifold can be done in several nonequivalent ways. In this work we will focus on the $b$-boundary 
method \cite{sch}. This method allows a classification of singularities in terms of parallel propagated frames, it distinguishes between points at 
infinity and points at a finite distance, and it generalises the idea of affine length to all curves regardless of them being geodesic or not. 
Other common techniques to attach boundaries to Lorentzian manifolds are conformal boundaries \cite{hawking1, confo} and the causal 
boundaries \cite{hawking1,flores2} which we describe below. In addition, we would like to mention other constructions such as the $a$-boundary \cite{scott,scott1} 
and boundaries constructed from light rays \cite{low}.

 The conformal boundary allows us to study the structure of the metric at ``infinity''. The idea of conformal compactification is to bring points 
 at ``infinity'' on a non-compact pseudo-Riemannian manifold $({\mathcal{M}},g)$ to a finite distance (in a new metric) by a conformal rescaling 
 of the metric ${\tilde g} = \Omega^2 g$. This precise definition of conformal compactification only applies to an asymptotically simple spacetime. $({\mathcal{M}},g)$
 is asymptotically simple, if there are another smooth Lorentz manifold and associated metric $(\tilde{{\mathcal{M}}},\tilde{g})$ such that:
 \begin{itemize}
   \item  ${\mathcal{M}}$ is an open sub-manifold of ${\tilde {\mathcal{M}}}$ with smooth boundary $\partial{{\mathcal{M}}}$ called the conformal boundary;
   \item there exists a smooth scalar field $\Omega$ on ${\tilde {\mathcal{M}}}$ such that $\tilde{g}=\Omega^{2}g$ on ${\mathcal{M}}$, and 
   $\Omega=0,d\Omega\neq0$ on $\partial{{\mathcal{M}}}$;
   \item every null geodesic in ${\mathcal{M}}$ acquires a future and past endpoints on $\partial{{\mathcal{M}}}$.
 \end{itemize}
 
This technique has the evident drawback that it can only be applied to this kind of spacetimes \cite{confo}. Moreover, 
notice that in Minkowski spacetime the conformal boundary is given by  $\partial{{\mathcal M}}=\mathscr{I}^-\cup \mathscr{I}^+$ (where $\mathscr{I}^-$ corresponds 
to null-past infinity and $\mathscr{I}^+$ corresponds to null-future infinity) while $i^{o},i^{+},i^{-}$ which correspond to spacelike infinity,
future timelike infinity, and past timelike infinity respectively do not belong to the conformal boundary (the thorough reader can
find in \cite{hawking1} formal definitions of $\mathscr{I}^{-}, \mathscr{I}^{+}, i^{o}, i^{+}, i^{-}$). The reason for this 
is because $\partial{{\mathcal M}}$ is not a smooth manifold at these points. Despite this, the conformal boundary has been successfully applied to study isolated systems in General Relativity \cite{confo} and to the AdS-CFT correspondence \cite{ads,marolfgrg}.
 
On the other hand, the causal boundary of a spacetime consists on attaching a boundary that depends only on the causal structure. However, this 
implies on this particular construction that one is not able to distinguish between boundary points and points at infinity. Moreover
one has to assume that $({\mathcal M},g)$ is strongly causal. This construction relies on indecomposable past sets (IP) and indecomposable
future sets (IF) which we now define. An open set $U$ is an IP if it satisfies $I^{-}(U)\subset U$ and cannot be expressed as the 
union of two proper-open subsets $V$ and $W$ satisfying $I^{-}(V)\subset V$ and $I^{-}(W)\subset W$ respectively. 
Similarly using $I^{+}$ one can define IF. The class of IPs can be divided into two classes: proper IPs (PIPs) which are 
of the form $I^{-}(p)$ for $p\in {\mathcal M}$, and terminal IPs (TIPs) which are not formed by the history of any point in ${\mathcal{M}}$.
We shall denote by $\overbrace{{\mathcal M}}$ the set of all IPs of the space $({\mathcal{M}},g)$ and $\underbrace{{\mathcal M}}$ the set of all IFs 
of the space $({\mathcal M},g)$. Originally one define a suitable topology on $\overbrace{{\mathcal M}},\underbrace{{\mathcal M}}$ to identify IFs and 
IPs and one can form a space ${\mathcal M}^{*}={\mathcal M}\bigcup \Delta$ where $\Delta$ is called the causal boundary \cite{hawking1}. However, this
topology presents some problems that have led to several redefinitions. A full revision of the causal boundary and its relationship 
with the conformal boundary can be found in \cite{flores2}. Also its relation with boundaries 
in Riemannian and Fislerian manifolds can be found in \cite{flores3}.

There is also the $b$-boundary. This is a method developed by Schmidt which allows one to attach a 
boundary $\partial\mathcal{M}$ called the \emph{$b$-boundary} to any incomplete spacetime $\mathcal{M}$ (or even to any manifold with a connection). 
The procedure consists on constructing a Riemannian metric for the frame bundle $L({\mathcal{M}})$ or the orthonormal bundle $O({\mathcal{M}})$ called 
the Schmidt metric. This metric is then used to generalise the idea of affine length to all curves. This generalisation is important because it 
helps to unify some elements of Riemannian geometry with Lorentzian geometry. For example, while only in the Riemannian case geodesic completeness 
implies that every curve is complete; the notion of $b$-completeness implies completeness of every curve in both signatures. The definition of a curve
we are using here is a piecewise-$C^{1}$ curve.

The structure of this paper is as follows. In Sec.\ \ref{pre} we give a general overview of the mathematical preliminaries needed. 
In section \ref{bb} we describe how the Schmidt metric and the $b$-boundary are constructed following the procedure by Schmidt \cite{sch}. In Sec.\ \ref{1+1} we discuss
the geometry of the orthonormal-bundles for $1+1$ conformally flat spacetimes
In the last section, namely Sec.\ \ref{dis}, we 
discuss the results in the context of gravitational singularities in general relativty.

%-------------------------------------------------------------------------------------------------------------------------------------------------------
% PRELIMINARIES SECTION
%-------------------------------------------------------------------------------------------------------------------------------------------------------

\section{Preliminaries} \label{pre}
As a first step, let us present some of the required concepts of differential geometry. We present the basic concept 
of fibre bundles, $G$-principal bundles, solder forms and  connections and 
The manifolds we consider in this paper are paracompact, $C^{\infty}$, connected, and Hausdorff.

%-----------------------------------------------------------------------------------------------
% FIBRE BUNDLES AND G-PRINCIPAL BUNDLES SUBSECTION
%-----------------------------------------------------------------------------------------------

\subsection{Fibre Bundles and $G$-Principal Bundles}
A {\emph{Fibre bundle}} with fibre $\mathcal{F}$ is a manifold ${E}$ with a surjetive map $\pi:{{E}}\rightarrow \mathcal{M}$ where there
is a neighbourhood $\mathcal{U}$ at each point $p$ of $\mathcal{M}$ such that $\pi^{-1}({\mathcal{U}})$ is isomorphic to ${\mathcal{U}}\times {\mathcal{F}}$, i.e., 
for each point $p\in {\mathcal{U}}$ there is a diffeomorphism $\phi_{p}$ of $\pi^{-1}(p)$ onto ${\mathcal{F}}$ such that the map 
$\psi(\overline{p})=(\pi(\overline{p}),\phi_{\pi(\overline{p})})$ is a diffeomorphism.  We call ${\mathcal{M}}$ the base space of the fibre bundle $E$.

A {\emph{$G$-principal bundle}} $P$ over a manifold ${\mathcal{M}}$ is a fibre bundle with fibre a lie group $G$ with a continuous right action $R_{g}$ that acts freely:
$$ (\overline{p},g) \in P\times G \rightarrow R_{g}(\overline{p})\in P$$ 
and satisfies that ${\mathcal{M}}$ is the quotient space of $P$ by the equivalence relation induced by $G$ \cite{differe}.

Let $\mathcal{M}$ be a $n$-dimensional manifold. A \emph{frame} $\{{\bf{E}}^{a}\}_{p}$ at point $p$ is an ordered 
basis of $T_{p}$. Let $L(\mathcal{M})$ be the set of all frames $\{{\bf{E}}^{a}\}$ at all points on $\mathcal{M}$ with the
projection $\pi$ sending a frame at $p$ to $p$.  Then the \emph{general linear group} $GL(n,\mathbb{R})$ has a natural
action on $\{{\bf{E}}^{a}\}$, {i.\ e., } given $(\{{\bf{E}}^{a}\},A_{a}^{b})$ the action of $A_{a}^{b}\in GL(n,\mathbb{R})$  
on $\{{\bf{E}}^{a}\}$ is $ \{{\bf{E}}^{b}={A_{a}^{b}\bf{E}}^{a}\}$. If $\{x^{a}\}$ are coordinates on $\mathcal{M}$ and we choose the 
frame $\{\frac{\partial}{\partial x^{a}}\}$, then it can be shown that the coordinates $(x^{a}, \beta^{{a}}_{{b}})$ are a local coordinate 
system of  ${L{\mathcal{M}}}$, where $\beta^{{a}}_{{b}}$ represent the $ab$ element for the change of basis matrix $\beta$ between $\{\frac{\partial}{\partial x^{a}}\}$ 
and any other frame  $\{{\bf{E}}^{b}\}$. In fact this choice makes $L({\mathcal{M}})$ a $G$-principal bundle called the \emph{frame bundle}.
Moreover, if we have a metric in $\mathcal{M}$ and we restrict the frames to just orthonormal frames, we obtain another $G$-principal bundle called 
the \emph{orthonormal frame bundle} $O(\mathcal{M})$. The associated Lie Group to $O(\mathcal{M})$ is then the \emph{orthonormal group}, $SO(n,\mathbb{R})$ 
or $SO^{+}(1,n,\mathbb{R})$. This last definition is signature dependent. 
 
Every tangent space $T_{\overline{p}}P$ of a $G$-principal bundle $P$ has a subspace called the \emph{vertical subspace} $V_{\overline{p}}$. 
This subspace is given by the kernel of the differential $D\pi$ restricted at $\overline{p}$. Explicitly, 
$$V_{\overline{p}}=\{{\bf{X}}\in T_{\overline{p}}P| D\pi_{\overline{p}}({\bf{X}})=0\in T_{\pi(p)}{\mathcal{M}} \}.$$        

%-----------------------------------------------------------------------------------------------
% SOLDER FORM AND CONNECTIONS SUBSECTION
%-----------------------------------------------------------------------------------------------

\subsection{Solder form and Connections}
The \emph{solder form} of a frame bundle $L(\mathcal{M})$ is the map:
$$\theta:TL({\mathcal{M}})\rightarrow \mathbb{R}^{n}:Q\rightarrow \overline{p}^{-1}(D\pi(Q))$$
where  $Q$ is an element of $T_{\overline{p}}L\mathcal{M}$ and $\overline{p}$ is a linear map from $\mathbb{R}^n$ to  $T_{\pi(\overline{p})}$ that sends the canonical vectors to a choice of basis on $T_{\pi(\overline{p})}$ . 
The solder form for the orthonormal bundle $O(\mathcal{M})$ is defined similarly. Notice that $V_{\overline{p}}\subset ker(\theta)$. 
  
A \emph{connection} $\overline{\nabla}$ on a $G$-principal bundle is an assignment of a subspace $H_{\overline{p}}$ called 
the \emph{horizontal subspace} of $T_{\overline{p}}(P)$ for all $\overline{p}$ in $P$ such that:
\begin{itemize}
  \item $T_{\overline{p}}P=H_{\overline{p}}\oplus V_{\overline{p}}$.
  \item $H_{\overline{p}g}=D_{\overline{p}}R_{g}(H_{\overline{p}})$ for every $\overline{p}\in P$ and $g\in G$. 
\end{itemize}
 
A \emph{connection form} $\varpi$ of a connection $\overline{\nabla}$ in a $G$-principal bundle is a $C^{\infty}$ map 
$$\varpi:TP\rightarrow \mathfrak{g}$$ 
with the following properties:
\begin{itemize}
  \item if $\varpi(X)=0$ then $X\in H_{\overline{p}}$ for some $\overline{p}$ in $P$,
  \item for all $g$ in $G$ and all $C^{\infty}$ maps $X:P\rightarrow TP$,  $\varpi(DR_{g}(X))=ad_{*}(g^{-1})\varpi(X)$, and
  \item for all $\vec{g}\in\mathfrak{g}, \varpi(X^{*}_{\overline{p}})=\vec{g}$ where $X^{*}_{p}$ is the tangent vector at $t=0$ of
  a curve given by $\gamma(t)=R_{\exp{t}\vec{g}}(\overline{p})$.
\end{itemize} 
Let us remind the reader that connections and connection forms uniquely determine one another. 

In coordinates the connection form $\varpi$ is written as $\varpi=\sum_{a,b}{\varpi}^{a}_{\;b}{\bf{F}}^{a}_{\;b}$ where
\begin{equation}\label{connection}
  {\varpi}^{a}_{\;b}=\sum_{c}\left((\beta^{-1})^{a}_{\;c} d\beta^{\;c}_{b}+\sum_{d,e}(\beta^{-1})^{a}_{\;c}\Gamma ^{c} _{\;de}\beta^{e}_{\;b}dx^{d}\right),
\end{equation}
where $(\beta^{-1})^{a}_{\;c}$ is the inverse of the matrix $\beta^{a}_{\;c}$ and $\Gamma^{{a}}_{\;{b}{c}}$ are the Christoffel symbols. 

The solder form $\theta$ is then given by $\theta=\sum_{a} {\theta}^{a}{\bf{e}}^{a}$ where
\begin{equation}\label{solder}
\theta^{a}=\sum_{c}(\beta^{-1})^{a}_{\;c}dx^{c}.
\end{equation}

and ${\bf{e}}^{a}$ is the natural basis of $\mathbb{R}^{n}$ \cite{differe}.

%-------------------------------------------------------------------------------------------------------------------------------------------------------
% THE SCHMIDT METRIC SECTION
%-------------------------------------------------------------------------------------------------------------------------------------------------------

\section{The Schmidt metric}\label{bb}
If one thinks of a singularity in classical Newtonian gravity, the statement that the gravitational field 
is singular at a certain location is unambiguous. As an example, take the gravitational potential of a spherical mass $M$ in Cartesian coordinates
$$V(t,x,y,z)=\frac{GM}{\sqrt{x^{2}+y^{2}+z^{2}}},$$
where $G$ is the gravitational constant, and the potential exhibits a singularity at the point $x=y=z=0$, for any time $t$ 
in $\mathbb{R}$. The location of the singularity is well defined because the coordinates have an intrinsic character which is independent of $V$.

However, in the case of GR the prescription given above can not work. This is due to the identification of the background 
spacetime with the gravitational field. Hence, only in the regions where the gravitational field is defined it is meaningful 
to talk about locations. Consider the spacetime with the line element
$$ds^{2}=-\frac{1}{t^{2}}dt^{2}+dx^{2}+dy^{2}+dz^{2},$$
defined on the manifold $\{(t,x,y,z)\in \mathbb{R}\backslash \{0\}\times \mathbb{R}^{3}\}$. If we say that there is a singularity 
at the point $t=0$, we will be speaking too soon for two reasons. The first one is that $t=0$ is not part of the manifold. 
It makes no sense to talk about $t=0$ as a location where the field diverges. The second thing is that the lack of an intrinsic 
meaning of the coordinates in GR must be taken seriously. By making the coordinate transformation $\eta=\log(t)$, we obtain the line element
$$ds^{2}=-d\eta^{2}+dx^{2}+dy^{2}+dz^{2},$$
on $\mathbb{R}^{4}$ which is an isometric extension of the previously defined spacetime. This spacetime is, of course, Minkowski spacetime 
which is non-singular \cite{hawking1}.

Another idea is trying to define a singularity in terms of invariant quantities such as invariant scalars. The reason for this 
is that if these quantities diverge then it matches our physical idea that objects must suffer stronger and stronger deformations 
as we encounter the singularity. These scalars are usually constructed from contractions of the Riemann tensor and its derivatives. 
Unfortunately, these scalars are not well-suited to define the complete geometry. Consider the metric
$$ds^{2}=dudv+H_{ij}(u)x^{i}x^{j}du^{2}-dx^{i}dx^{i},$$
given in the coordinates $(u,v,x^{1},x^{2})$ and where $H(u)$ is $C^{1}$.

This spacetime is known as a $pp$-wave spacetime and it can be shown that every polynomial curvature-scalar vanishes, 
despite the fact that in general the spacetime is not flat \cite{gibbons}.

A more troublesome feature of using scalars for defining singularities is that they are `too local' in the sense that they 
are evaluated at given points. Therefore, if the point is removed, the scalar cannot be computed directly  and we need an approximation procedure.

A precise mathematical way to approximate the ``missing points'' is to use convergent sequences of points on the manifold.
In this case the formal statement is: ``The sequence $\{R(x_{n})\}$ diverges while the sequence $\{x_{n}\}$ converges to $y$'', where $R(x_{n})$ is some 
scalar curvature invariant evaluated at $x_{n}$ in $\mathcal{M}$ and $y$ is some point not necessarily in $\mathcal{M}$.

In Riemannian geometry, the notion of distance allows us to define Cauchy sequences $\{x_{n}\}$ and therefore a notion of convergence. 
Moreover, if every Cauchy sequence converges in $\mathcal{M}$ then every geodesic can be extended indefinitely. This means we can take 
the domain of every geodesic to be $\mathbb{R}$. In this case, we say that $\mathcal{M}$ is \emph{geodesically complete}. In fact, 
the converse is also true: if $\mathcal{M}$ is geodesically complete then $\mathcal{M}$ is metrically complete, {\it i.\ e.,} every Cauchy 
sequence converges to a point in $\mathcal{M}$ \cite{palmas}. This allows us to use Cauchy sequences or sequences of points along geodesics 
as our sequences of points.

The Riemannian case is a useful example, but as soon as we move to Lorentzian geometry, which we take as the correct 
geometrical setting for GR, the previous discussion cannot be used as stated. The reason is that Lorentzian metrics do 
not have a distance function defined and therefore Cauchy sequences cannot be defined. Thus, one is restricted to the 
notion of geodesically complete manifolds in the Lorentzian case.

Moreover, the existence of three kinds of vectors available in any Lorentzian metric defines three nonequivalent notions of 
geodesic completeness ---depending on the character of the tangent vector of the curve--- spacelike completeness, null 
completeness and timelike completeness, which are, unfortunately, not equivalent. It is possible to construct spacetimes 
with the following characteristics \cite{kundt, geroch, beem}:
 \begin{itemize}
   \item timelike complete, spacelike and null incomplete,
   \item spacelike complete, timelike and null incomplete,
   \item null complete, timelike and spacelike incomplete,
   \item timelike and null complete, spacelike incomplete,
   \item spacelike and null complete, timelike incomplete, or
   \item timelike and spacelike complete, null incomplete.
 \end{itemize}
Furthermore, there are examples of a geodesically null, timelike and spacelike complete spacetimes with an inextendible 
timelike curve of finite length \cite{geroch, beem}. A particle following this trajectory will experience bounded acceleration 
and in a finite amount of proper time its spacetime location would stop being represented as a point in the manifold.

In order to overcome this, Schmidt provided an elegant way to generalise the idea of affine length to all curves, regardless 
of such curves being geodesic or not. This construction in the case of incomplete curves allows to attach to the spacetime $\mathcal{M}$ 
a topological boundary $\partial\mathcal{M}$ called the {$b$-boundary}. The procedure for constructing the Schmidt metric consists 
in building a Riemannian metric in the frame bundle $L({\mathcal{M}})$. We use the solder form $\theta$ on $L(\mathcal{M})$ and the
connection form $\varpi$ on $L(\mathcal{M})$ associated to the Levi-Civita connection $\nabla$ on $\mathcal{M}$ to do this. Explicitly, the Schmidt metric is given by
\begin{equation}
  \overline{g}({\bf{X}},{\bf{Y}})= \theta({\bf{X}})  \cdot \theta({\bf{Y}})+\varpi({\bf{X}})\bullet  \varpi({\bf{Y}}),
\end{equation}
where ${\bf{X}},{\bf{Y}}\in T_{\overline{p}}P$ and $\cdot,\bullet$ are the inner products in $\mathbb{R}^{n}$ 
and $\mathfrak{g}\cong\mathbb{R}^{n^{2}}$ respectively.  { The construction of the Schmidt metric is more general and can be applied to any manifold with a connection. This connection does not necessarily need to ber a metric compatible connection \cite{sch}. However, as mentioned above, we will use the Levi-Civita connection because we will always assume a metric on the manifold. }

Let $\gamma:[a,b]\rightarrow {\mathcal{M}}$ be a piecewise-$C^{1}$ curve through $p$ in $\mathcal{M}$. A curve $\overline{\gamma}:[a,b]\rightarrow L({\mathcal{M}})$  in $L({\mathcal{M}})$ is called the \emph{lift of the curve } $\gamma$ if it satisfies $\pi(\overline{\gamma})=\gamma$ and $D\pi(\dot{\overline{\gamma}})=\dot{\gamma}$. The length of $\overline{\gamma}$ with respect to the Schmidt metric, is
$$L_{\overline{\gamma}}(b)=\int^{b}_{a}\|\dot{\overline{\gamma}(\eta)} \|_{\overline{g}} d\eta,$$
which is called the \emph{generalised affine-length} of $\gamma$. We can then use this to 
re-parametrise $\gamma$ which generalises the notion of an affine parameter. 
In the case where $\gamma$ is a geodesic parametrised by $L_{\overline{\gamma(t)}}$, it is 
parametrised with respect to an affine parameter. If every curve in a spacetime $\mathcal{M}$ with finite 
generalised-affine-length has endpoints, we call this spacetime \emph{b-complete}. If it is not $b$-complete we 
say that the spacetime is \emph{b-incomplete.}

Notice that if there is a curve $\gamma$ in $\mathcal{M}$ that has finite affine-length and no endpoint then the 
lift curve $\overline{\gamma}$ cannot have an endpoint. Otherwise, if $\overline{p}$ is the endpoint of $\overline{\gamma}$, $\pi(\overline{p})=p$ 
would be an endpoint of $\gamma$ contradicting the incompleteness of $\gamma$. The previous remark shows that geodesic incompleteness 
implies $b$-incompleteness. The converse is not true as Geroch's example \cite{geroch} shows a $b$-incomplete spacetime that is geodesically complete.
Therefore $b$-incompleteness is a generalisation of geodesic incompleteness.

Now given an incomplete spacetime $\mathcal{M}$, using the Riemannian metric $\overline{g}$ on $L({\mathcal{M}})$, we can
`Cauchy complete' $L({\mathcal{M}})$. Let us denote by $\overline{L(\mathcal{M})}$ the Cauchy completion of $L({\mathcal{M}})$.

We define the quotient space $\overline{\mathcal{M}}= \overline{L(\mathcal{M})}/G^{+}$, where $G^{+}$ is the connected component of 
the identity of $GL(n;\mathbb{R})$ under the equivalence of orbits, { i.e., } 
$(\overline{p},g)\in\overline{L(\mathcal{M})}\sim (\overline{q},g')\in\overline{L(\mathcal{M})}$ if $\overline{p}=\overline{q}$ and 
there is $h\in GL(n;\mathbb{R})$ such that $g=hg'$. 
This quotient induces a topology in $\overline{\mathcal{M}}$ by taking the finest topology that makes 
the map $\pi:\overline{L(\mathcal{M})}\rightarrow \overline{\mathcal{M}}$ continuous and therefore $\overline{\mathcal{M}}$ is a topological space.
However, it does not imply that $\overline{\mathcal{M}}$ is a manifold. Finally we 
can characterise the $b$-boundary as the set $\partial\mathcal{M}=\overline{\mathcal{M}}\backslash {\mathcal{M}}$.

We repeat the same construction for subgroups of $GL(n;\mathbb{R})$. In particular, a common choice in the Lorentzian case
is the subgroup of all Lorentz transformations preserving both orientation and direction of time, which is called the proper
orthochronous Lorentz group and it is denoted by $SO^{+}(1, n)$. In a completely analogous way we can form the
quotient $\overline{{\mathcal{M}}}= \overline{O({\mathcal{M}})}/SO^{+}(1,n;\mathbb{R})$ and define the $b$-boundary as
the set $\partial\mathcal{M}=\overline{{\mathcal{M}}}\backslash {\mathcal{M}}$. The completion using $SO^{+}(1, n)$ is homeomorphic to the completion using $GL(n;\mathbb{R})$ \cite{amores2}.
The advantage of this construction is that $O(\mathcal{M})$ is 
a manifold of dimension $n+\frac{n(n-1)}{2}$ instead of the $n+n^{2}$ dimensions of $L({\mathcal{M}})$. Also, the construction can
be carried in a manifold with a Riemannian metric, in that case $ \overline{{\mathcal{M}}}$ is homeomorphic to the Cauchy completion.
of $\mathcal{M}$ \cite{clarke}. This reinforces the conviction that the $b$-boundary is a natural way to attach boundaries to manifolds with connections.

%-------------------------------------------------------------------------------------------------------------------------------------------------------
% THE b-BOUNDARY FOR 1+1 SPACETIMES SECTION
%-------------------------------------------------------------------------------------------------------------------------------------------------------

\section{The Schmidt metric of 1+1 spacetimes }\label{1+1}

In this section, we locally construct the Schmidt metric for general $1$+$1$ spacetimes.  Moreover, we find a relationship between the scalar curvature of the Schmidt metric on ($O({\mathcal{M}})$, $\overline{g}$) and the scalar curvature of (${\mathcal{M}}$, ${g}$). Finally, we give several explicit examples.

{\bf{Notation:}}We use overlines to denote the Riemannian geometric quantities that belongs to $O({\mathcal{M}})$ while 
the geometric quantities without any overline belong to the Lorentzian manifold ${\mathcal{M}}$.

%-----------------------------------------------------------------------------------------------
% THE SCHMIDT METRIC FOR 1+1 CONFORMAL SPACETIMES
% SUBSECTION
%-----------------------------------------------------------------------------------------------

\subsection{The Schmidt metric for 1+1 conformal spacetimes}    
Let $\mathcal{M}$ be a 2-D manifold with a Lorentzian metric $g$ and an orthonormal bundle $O({\mathcal{M}})$. Then, 
we can find coordinates $(v,w)$ which locally transform the line element of the metric $g$ to the following form \cite{isothermal}:
\begin{equation}\label{conformal}
  ds^{2}=\Omega^{2}(v,w)(-dv^{2}+dw^{2}).
\end{equation} 
An orthonormal basis is then given by the vector fields
\begin{align}
 E_{1}=&\frac{1}{\Omega}\frac{\partial}{\partial{v}}\quad {\rm and} \\
  E_{2}=&\frac{1}{\Omega}\frac{\partial}{\partial{w}}. 
\end{align}

The orthonormal basis prescribed above is not unique. Any other orthonormal basis is of the form  
\begin{eqnarray}
  \tilde{E}_{1}&=&\cosh\chi \frac{1}{\Omega}\frac{\partial}{\partial{v}}+\sinh\chi \frac{1}{\Omega}\frac{\partial}{\partial{w}}\\
\tilde{E}_{2}&=&\cosh\chi \frac{1}{\Omega}\frac{\partial}{\partial{w}}+\sinh\chi\frac{1}{\Omega}\frac{\partial}{\partial{v}}
\end{eqnarray}
for some $\chi\in\mathbb{R}$.

Let us notice that the coefficients of such a basis with respect to $\frac{\partial}{\partial v},\frac{\partial}{\partial w}$ define a unique  non-singular matrix $\beta $ with inverse $\beta^{-1}$: 
\begin{align}
 \beta =\frac{1}{\Omega}
 \begin{pmatrix}
\cosh\chi & \sinh\chi\\
\sinh\chi & \cosh\chi \\
   \end{pmatrix}, \quad 
{\rm and }\quad  \beta^{-1}={\Omega}
   \begin{pmatrix}
\cosh\chi & -\sinh\chi\\
-\sinh\chi & \cosh\chi \\
 \end{pmatrix}. 
\end{align}
These matrices are important in the sense that they are useful to define local coordinates on $O({\mathcal{M}})$ as follows:
\begin{align}
\big\{ v,w,& \frac{1}{\Omega}\left(\cosh\chi {\frac{\partial}{\partial{v}}}+\sinh\chi \frac{\partial}{\partial{w}}\right), \nonumber \\
& \frac{1}{\Omega}\left(\cosh\chi \frac{\partial}{\partial{w}}+\sinh\chi{\frac{\partial}{\partial{v}}}\right)| (v,w)\in \mathcal{M}, \chi\in\mathbb{R}  \big\}.   
\end{align}

As stated in section \ref{bb}, the Schmidt metric $\overline{g}$ for any ${\bf{X}},{\bf{Y}}\in TO(\mathcal{M})$ on  $O({\mathcal{M}})$ is given by 
\begin{equation}\label{Schmidtmetric}
\overline{g}({\bf{X}},{\bf{Y}}):\varpi({\bf{X}})\cdot\varpi({\bf{Y}})+\theta({\bf{X}})\cdot\theta({\bf{Y}})  
\end{equation}
where $\varpi$ is the connection form on $O({\mathcal{M}})$ and $\theta$ the solder form.

Now let us consider a curve $\overline{\gamma}(s)$ in $O(\mathcal{M})$ given by $\overline{\gamma}:s\in[a,b]\mapsto (v(s),w(s),\beta^{a}_{\;c}(s))$ and evaluate $\theta(\dot{\overline{\gamma}})$ and $\varpi(\dot{\overline{\gamma}})$. Explicitly we have
\begin{align}
\theta(\dot{\overline{\gamma}}) =  
\Omega \begin{pmatrix}
\dot{v}\cosh\chi-\dot{w}\sinh\chi\\
-\dot{v}\sinh\chi+\dot{w}\cosh\chi \\
 \end{pmatrix},
\end{align} 
 and 
\begin{align}
{\varpi}(\dot{\overline{\gamma}}) = 
 \begin{pmatrix}
0 & \dot{\chi}+\frac{1}{\Omega}((\partial_{v}{\Omega})\dot{w}+(\partial_{w}{\Omega})\dot{v})\\
\dot{\chi}+\frac{1}{\Omega}((\partial_{v}{\Omega})\dot{w}+(\partial_{w}{\Omega})\dot{v})& 0 \\
 \end{pmatrix},
\end{align} 
%\noindent
where we have used \ref{connection} and \ref{solder}. Then, the line element for the Schmidt metric using a general inner product can be written as:
\begin{align}\label{smgeneral}
ds^{2} = & \Omega^{2}(v,w)[(a_{11} \cosh^2\chi+a_{22} \sinh^2\chi)dv^2-2(a_{11}+a_{22})\sinh\chi\cosh\chi dvdw \nonumber \\
&+(a_{22} \cosh^2\chi+a_{11} \sinh^2\chi)dw^2+2 a_{12}(\cosh2\chi dvdw-\sinh2\chi(dv^2+dw^2))] \nonumber \\
&+(b_{22}+2b_{24}+b_{44})\left( d\chi+\frac{1}{\Omega}\left(\frac{\partial\Omega}{\partial v}dw+\frac{\partial\Omega}{\partial w}dv\right)\right)^{2}.
\end{align}

where $A=(a_{ij}), B=(b_{ij})$ are symmetric matrixes with positive eigenvalues. 

%\begin{align}
 %A = \begin{pmatrix}
%a_{11} & a_{12}\\
%a_{21} & a_{11} \\
 %  \end{pmatrix}
%{\rm and }\quad  B=
 %\begin{pmatrix}
%b_{11} & b_{12} & b_{13} & b_{14}\\
%b_{21} & b_{22} & b_{23} & b_{24}\\
%b_{31} & b_{32} & b_{33} & b_{34}\\
%b_{41} & b_{42} & b_{43} & b_{44}\\
% \end{pmatrix},
%\end{align} 

It can be shown that using two different inner products produce two unifomly equivalent metrics \cite{dodson}. 

In application it is commonly used the Euclidean innner product which give the line element for the Schmidt:
\begin{align}\label{sm}
  ds^{2}= \Omega^{2}(v,w)(\cosh(2\chi) (dv^{2}+dw^{2})-&2\sinh(2\chi) dv dw)\\
   +&\left( d\chi+\frac{1}{\Omega}\left(\frac{\partial\Omega}{\partial v}dw+\frac{\partial\Omega}{\partial w}dv\right)\right)^{2}.\nonumber
\end{align}

We avoid quoting long tensorial expressions for the curvature tensors and give only the result for the Ricci scalar of \eqref{sm} in terms of $\Omega$ and its derivatives, but we have that $\overline R$  is given by 

\begin{align}\label{prericci}
\overline{R}=-\frac{1}{2 \Omega^8}\left(\left(\Omega_{ww}-\Omega_{vv} \right) \Omega- \left(\Omega_w^2-\Omega_v^2\right)\right)^{2}-2.
\end{align}

Taking into account that 

\begin{align}
R=-\frac{2}{\Omega^4}((\Omega_{ww}-\Omega_{vv})\Omega-(\Omega_w^2-\Omega_v^2))
\end{align}
This means that Eq. (\ref{prericci}) becomes 
\begin{align}\label{ricci}
{\bar R}=-\frac{1}{8}R^2-2
\end{align}

Notice that \eqref{ricci} is the relationship between both scalar curvatures. As direct consequences we can establish the negativity of the  Ricci scalar for any Schmidt metric in the Lorentzian signature. In \ref{s1} we give counterexamples that such a condition does not hold in the Riemannian case. {Also,  Eq. \ref{ricci} has been obtained using the Levi-Civita connection. Therefore, using another connection, even in the Lorentzian case, may not hold.}

Now we calculate such scalar curvatures for some physical spacetimes.

%\begin{align}
%4 \Omega^8 K(v,w)=\left[4 \text{sech}(2 \chi ) \left(\partial_{ww}\Omega-\partial_{vv}\Omega\right) \Omega^5+4 \text{sech}(2 \chi ) \left((\partial_v\Omega )^2-(\partial_w\Omega )^2\right) \Omega^4 \right. \\
%\left. +\left(\partial_{ww}\Omega-\partial_{vv}\Omega\right)^2     
%\Omega^2-2 \left((\partial_w\Omega)^2-\partial_v\Omega )^2\right) \left(\partial_{ww}\Omega
%-\partial_{vv}\Omega \right) \Omega+\left((\partial_w\Omega)^2-(\partial_v\Omega )^2\right)^2-4 \Omega^8\right]
%\end{align}
 
%ne again with help form computer algebra we obtain analytical expressions for the Cotton-York tensor as it was defined in Eq.\ \eqref{CY}. These general expressions include 4-th order derivatives of $\Omega(v,w)$ which are too long to write here. With the explicit components of the Cotton-York we calculate the corresponding eigenvalues. Those eigenvalues are then used to classify each spacetime according to Sec.\ \ref{sec:CY} \cite{garcia}.

%-----------------------------------------------------------------------------------------------
% THE SCHMIDT METRIC FOR MINKOWSKI SPACETIME SUBSECTION
%-----------------------------------------------------------------------------------------------

\subsection{The Schmidt metric of Minkowski spacetime}\label{m}
We can write the Schmidt metric in the form
\begin{equation}
ds^{2}=(dt^{2}+dx^{2})\cosh(2\chi)-2dtdx\sinh(2\chi)+d\chi^{2}.  
\end{equation}

Now let us consider the change of coordinates: $t=u+\tilde{v}, x=u-\tilde{v}$ and write:
\begin{equation}
ds^{2}=2(\cosh2\chi+\sinh2\chi)du^{2}+2(\cosh2\chi-\sinh2\chi)d\tilde{v}^{2}+d\chi ^{2}
\end{equation}
or in an equivalent manner
\begin{equation}
  ds^{2}=2e^{2\chi}du^{2}+2e^{-2\chi}d\tilde{v}^{2}+d\chi^{2}.
\end{equation}

%We are going to use now Cartan's method \cite{differe} to calculate the curvature. First let us choose as an orthonormal co-frame defined by
%\begin{align}
%l^{0}=&{{\sqrt{2}}}e^{\chi}du,\nonumber\\
%l^{1}=&{{\sqrt{2}}}e^{-\chi}d\tilde{v}, \quad {\rm and}\nonumber\\
%l^{2}=&d\chi,  
%\end{align}
%where the corresponding one-forms are
%\begin{align}\label{d00}
% dl^{0}=&{{\sqrt{2}}}e^{\chi}d\chi\wedge du=-l^{0}\wedge l^{2},\nonumber\\
 %dl^{1}=&-{{\sqrt{2}}}e^{-\chi}d\chi\wedge d\tilde{v}=l^{1}\wedge l^{2}, \quad {\rm and}\nonumber\\
 %dl^{2}=&0.
%\end{align}

%Eq.\ \eqref{cartan} allows us to write \eqref{d00} in terms of connection one-forms:
%\begin{align*}
%\label{d0}
%  dl^{0}&=-l^{0}\wedge l^{2} \nonumber \\
 %       &= -\varpi^{0}_{\;0}\wedge l^{0}-\varpi^{0}_{\;1}\wedge l^{1}-\varpi^{0}_{\;2}\wedge l^{2},  \\  
 %   dl^{1}&=-l^{1}\wedge l^{2}\nonumber \\
  %      &= -\varpi^{1}_{\;0}\wedge l^{0}-\varpi^{1}_{\;1}\wedge l^{1}-\varpi^{1}_{\;2}\wedge l^{2}, \quad{\rm and}     \nonumber \\
  %dl^{2}&=0\nonumber \\
   %     &= -\varpi^{2}_{0}\wedge l^{0}-\varpi^{2}_{\;1}\wedge l^{1}-\varpi^{2}_{\;2}\wedge l^{2}.     \nonumber 
%\end{align*}
%Then we have
%\begin{align*}
 %\varpi^{0}_{\;2}=&l^{0}=-\varpi^{2}_{\;0}, \nonumber\\
 %\varpi^{1}_{\;2}=&-l^{1}=-\varpi^{2}_{\;1}\nonumber
%\end{align*}
%and all other components are zero.

%Using Eqs.\ \eqref{d0} and \eqref{cartan} we obtain:
%\begin{align}
%\Omega^{0}_{\;2}=&-l^{0}\wedge l^{2}\nonumber\\
%\Omega^{0}_{\;1}=&l^{0}\wedge l^{1}\nonumber\\
%\Omega^{1}_{\;2}=&-l^{1}\wedge l^{2}. \nonumber
%\end{align}
 
 We explicitly calculate $\overline{R}_{ab}$ and get 
\begin{equation}
  \overline{R}_{\chi\chi}=-2,
\end{equation}
and all other components are zero. The Ricci scalar is then
\begin{equation}
 \overline{R}=-{2}.
\end{equation}
Hence, the geometry in the bundle is not flat even if Minkowski spacetime is flat.

%The sectional curvature is given by 
%\begin{align*}
% K(u,v)=-1\\
%\end{align*}

%-----------------------------------------------------------------------------------------------
% THE SCHMIDT METRIC FOR MINKOWSKI SPACETIME SUBSECTION
%-----------------------------------------------------------------------------------------------

%----------------------------------------------------------------------------------------
% CASE MINKOWSKI: $\Omega (\nu, \omega) = 1$
%----------------------------------------------------------------------------------------

%{\color{magenta} The components of the Cotton-York Tensor are:
%\begin{align*}
%C_{\chi}^{\; i}=, \qquad C_{a}^{\; i}=,
%\end{align*}
%where $a={t, x}$ only.}

%-----------------------------------------------------------------------------------------------
% THE SCHMIDT METRIC FOR FRW SPACETIME SUBSECTION
%-----------------------------------------------------------------------------------------------

\subsection{The Schmidt metric of Friedmann-Robertson-Walker (FRW) spacetime}

%----------------------------------------------------------------------------------------
% CASE $\Omega^{2}=\eta^{q}$
%----------------------------------------------------------------------------------------

For simplicity, let us consider the case of the $1+1$ FRW cosmological model which can be obtained from the $4$-dimensional one by collapsing two spatial coordinates. This is equivalent to considering the injection map 
\begin{equation}
    h:(\eta,x)\rightarrow (\eta,x, y_{0},z_{0}):{\mathcal{M}}\rightarrow {\mathcal{N}}={\mathcal{M}}\times \Sigma,
  \end{equation}
where $\Sigma$ is a suitable two dimensional manifold. This way the four dimensional metric reduces to
\begin{equation}\label{metricafrw2}
ds^{2}=\eta^{q}(-d\eta^{2} +dx^{2}),
\end{equation}
for any value of $q>0$. The Ricci scalar corresponding to the spacetime described by \eqref{metricafrw2} is
\begin{equation}\label{metricafrw229}
R= -q \eta^{-2-q}.
\end{equation}

From Eq.\ \eqref{sm} the Schmidt metric in $O({\mathcal{M}})$, for our case study, takes the form
\begin{align}\label{sm1}
ds^{2}=\eta^{q}\left(\cosh(2\chi) (d\eta^{2}+dx^{2})-{2}\sinh(2\chi) d\eta dx\right)+\left(d\chi+\frac{q}{\eta} dx\right)^{2}.
\end{align}

Using computer algebra we calculated the Ricci tensor for the line element \eqref{sm1}. In components it is given by
\begin{align*}
\overline{R}_{\eta \eta} =& \frac{1}{8} q \eta^{-4 - q} (4 \eta^{2 + q} - q \cosh(2 \chi)),\\
\overline{R}_{x x} =& \frac{1}{32} q \eta^{-2 (3 + q)} (q^{3} - 16 \eta^{4 + 2 q} - 16 q \eta^{4 + 2 q} - 4 q (5 + 2 q) \eta^{2 + q} \cosh(2 \chi)),\\
\overline{R}_{\chi \chi} =& -2 + \frac{1}{8} q^{2} \eta^{-2 (2 + q)},\\
\overline{R}_{\eta x} =& \frac{1}{8} q^{2} (3 + q) \eta^{-4 - q} \sinh(2 \chi),\\
\overline{R}_{\eta \chi} =& \frac{1}{4} q (2 + q) \eta^{-3 - q} \sinh(2 \chi),\\
\overline{R}_{x \chi} =& \frac{1}{16} q \eta^{-5 - 2 q} (q^{2} - 16 \eta^{4 + 2 q} - 4 (2 + q) \eta^{2 + q} \cosh(2 \chi)).
\end{align*}
And the Ricci scalar is
\begin{equation}\label{ricciq}
\overline{R}= -2 - \frac{1}{8} q^{2} \eta^{-2 (2 + q)},
\end{equation}
which can equivalently be obtained from Eq.\ \eqref{ricci}.

%The sectional curvture is given by
%\begin{align*}
% K(\eta,x)=\frac{1}{2} q \eta^{-2 (q+2)} \text{sech}(2 \chi ) \left(\eta^{q+2}+\frac{q \text{sech}(2 \chi )}{8-8 \tanh^2(2 \chi )}\right)-1\\
%\end{align*}

%-----------------------------------------------------------------------------------------------
% THE SCHMIDT METRIC FOR DE SITTER SPACETIME SUBSECTION
%-----------------------------------------------------------------------------------------------

\subsection{The Schmidt metric of De Sitter and Anti-De Sitter spacetimes}

%----------------------------------------------------------------------------------------
% CASE DE SITTER
%----------------------------------------------------------------------------------------

Let us now consider the De Sitter and Anti-De Sitter models and study the behaviour of the corresponding curvature scalars. First, consider the De Sitter case. The two dimensional De Sitter spacetime for closed spatial sections is defined with the line element 
\begin{equation*}
 ds^{2} = - d \tau^{2} + \alpha^{-2} \cosh^{2} (\alpha \tau) d \omega^{2}.
\end{equation*}
 
To obtain the conformal form we make the change $\tan(\eta /2) = \tanh(\alpha \tau / 2)$, which leads to
\begin{equation}
 \label{desitter}
 ds^{2} = \frac{1}{\alpha^{2} \cos^{2}(\eta)} (-d \eta^{2} + d \omega^{2}).
\end{equation}
%\begin{equation*}
%ds^{2}= \frac{\alpha^{2}}{\cos^{2}(\eta)}(-d \eta^{2} + d \omega^{2} + d \nu^{2} + d \kappa^{2}).
%\end{equation*}
In these coordinates $(\eta, \omega)$ De Sitter space is conformal to the static Einstein universe \cite{hawking1}. The Ricci scalar for \eqref{desitter} is then
\begin{equation}\label{metricafrw224}
R = 2 \alpha^{2}.
\end{equation}

Using Eq.\ \eqref{sm} we get
\begin{equation}\label{DeSitterSM}
ds^{2} = \frac{1}{\alpha^{2} \cos^{2}(\eta)}(\cosh(2 \chi) (d \eta^{2} + d \omega^{2}) - 2 \sinh(2 \chi) d \eta d \omega) + (d \chi - \tan (\eta) d \omega)^{2}.
\end{equation}

The Ricci tensor is computed by taking $\Omega=\frac{1}{\alpha \cos(\eta)}$. The non-vanishing components are:
\begin{align*}
\overline{R}_{\eta \eta} & = -(1 + \frac{\alpha^{2}}{2} \cosh(2 \chi)) \sec^{2}(\eta),\\
\overline{R}_{\omega \omega} & = \frac{\alpha^{4}}{4}(1 + (-1 + 4 \alpha^{-4}) \cos(2 \eta) - 2 \alpha^{-2} \cosh(2 \chi)) \sec^{2}(\eta),\\
\overline{R}_{\chi \chi} & = -2 + \frac{\alpha^{4}}{2},\\
\overline{R}_{\eta \omega} & = \alpha^{2}\cosh(\chi) \sec^{2}(\eta) \sinh(\chi), \\
\overline{R}_{\omega \chi} & = (-2 + \frac{\alpha^{4}}{2}) \tan(\eta).\qquad
\end{align*}
Thus the Ricci scalar is%\footnote{{\color{red}{For the open spatial section it was obtained the same result for $\overline{R}$ and the flat spatial section is equivalent to FRW in the case of dark-energy domain.}}}
\begin{equation*}
\overline{R}= -2 - \frac{\alpha^{4}}{2}.
\end{equation*}
Notice that in the limit as $\alpha\to 0$ we recover the Minkowski limit once again.

%\begin{multicols}{2}
%\end{multicols}
%{\color{magenta} The components of the Cotton-York Tensor are:
%\begin{align*}
%C_{\chi}^{\; i}=&\frac{e^{-2\chi}\sec^2\eta(4\alpha^4-3)}{\alpha^2}\\
%C_{\eta}^{\; i}=&\frac{e^{-2\chi}\sec^2\eta (4\alpha^6-e^{2\chi}-5\alpha^2)(1+\tan\eta)}{2\alpha^4},\\
%C_{\omega}^{\; i}=& \frac{(\sec^2\eta (1 + \tan\eta + \alpha^2 e^{-2 \chi} (4 \alpha^4-5 + (12 \alpha^4-11) \tan\eta)))}{2 \alpha^4} .
%\end{align*}}

%----------------------------------------------------------------------------------------
% CASE ANTI-DE SITTER
%----------------------------------------------------------------------------------------

Now let us look at the Anti-De Sitter spacetime. The two dimensional Anti-De Sitter metric has the line element
\begin{equation*}
ds^{2} = \frac{1}{{\alpha^{2}} y^{2}}(-dt^{2} + dy^{2})
\end{equation*}
with $y>0$. The Ricci scalar is
\begin{equation}\label{metricafrw225}
R = -2 \alpha^{2}.
\end{equation}

We identify $\Omega=\frac{1}{\alpha y}$ and using Eq.\ \eqref{sm} we obtain
\begin{equation*}
ds^{2} = \frac{1}{\alpha^{2} y^{2}}(\cosh(2 \chi) (dt^{2} + dy^{2}) - 2 \sinh(2 \chi) dt dy) + (d \chi - \frac{1}{y} dt)^{2}.
\end{equation*}
Where the non-vanishing components of the Ricci tensor are
\begin{align*}
\overline{R}_{tt} = \frac{-2 + \alpha^4 - \alpha^2 \cosh(2 \chi)}{2 y^{2}},\quad
\overline{R}_{yy} = -\frac{2 + \alpha^{2} \cosh(2 \chi)}{2 y^{2}},\quad
\overline{R}_{\chi \chi} = -2 + \frac{\alpha^{4}}{2},\\
\overline{R}_{ty} = \frac{\alpha^{2} \cosh(\chi) \sinh(\chi)}{y^{2}},\quad
\overline{R}_{t \chi} = -\frac{-4 + \alpha^{4}}{2 y},\qquad\qquad
\end{align*}
and its trace is given by
\begin{equation*}
\overline{R}= -2 - \frac{\alpha^{4}}{2}.
\end{equation*}

{   Notice that for spacetimes that behave asymptotically as Anti-De Sitter spacetime, the curvature would behave similarly as in the Anti-De Sitter case as one approaches the asymptotic region. Moreover, in many applications such as in the AdS/CFT correspondance one uses a conformal compactification. In those cases it is neccesary to compute the curvature again because the curvature is not a conformal invariant.}
%\end{multicols}

%{\color{magenta} The components of the Cotton-York Tensor are:
%\begin{align*}
%C_{\chi}^{\; i}=&\frac{e^{-2\chi}(13+3y)}{y^2},\\
%C_{t}^{\; i}=&\frac{y-1 + 3 e^{2 \chi} (y-1) (4 + y) + 
% e^{-2 \chi} ( y (2 + 3 y)-37)}{2 y^3},\\
%C_{y}^{\; i}=& -\frac{(y-1 ) \cosh\chi (\cosh\chi + (23 + 6 y) \sinh\chi)}{y^3}.
%\end{align*}}

%-------------------------------------------------------------------------------------------------------------------------------------------------------
% DISCUSSION SECTION
%-------------------------------------------------------------------------------------------------------------------------------------------------------

\section{Discussion}\label{dis}

%\section{The topology of $\overline{{\mathcal{M}}}$}\label{ose}
In our exposition,  we obtained Eq.\ \eqref{sm} which is the line element of the 
Schmidt metric for all $1+1$ Lorentzian manifolds ${\mathcal{M}}$. This line element determines, via the curvature, 
all the local isometric invariants. If $\partial {\mathcal M}=\emptyset$, then the $3$-manifold corresponds to the orthonormal 
bundle where the fibres of the bundle are $SO^{+}(1,1)\cong \mathbb{R}$. Therefore, $O({\mathcal M})$ is not compact. 
If $\partial {\mathcal M}\neq\emptyset$, then the $3$-manifold is not necessarily a $G$-bundle (the group may not act freely or transitively). 
In fact there are general geometric conditions on the curvature to guarantee that the fibres above a boundary point are degenerate \cite{Sta}. This is, for example, the case when ${\mathcal{M}}$ is the Friedmann-Robertson-Walker spacetime. Then $\overline{O(\mathcal{M})}$ is not a 
$G$-bundle as the fibre over the singularity is a point instead of a copy of $SO(1,1)^{+}({\mathcal M} )$.
 { Moreover,  if there is a singularity not only in the past but also in the future both singularities are identified as the same boundary point\cite{clarke}. The degeneracy of the fibre} also affects the topology of $\overline{{\mathcal{M}}}$, which in the Friedmann-Robertson-Walker case is no longer Hausdorff. { In fact this topological behaviour is expected in general spacetimes when the fibre totally degenerates such as in Schwarzschild and Kasner metrics \cite{Sta}. }
However, there has been mathematical developments which allows to circumvent this undesirable situation by taking a canonical 
minimum refinement of the topology in the completion $M$ which $T2$-separates the spacetime
$M$ and its boundary $\partial M$ \cite{flores1}. { Notice that this result does not guarantee that points which one may consider physically different such as the initial and final singularity in a closed Friedmann-Robertson-Walker scenarios are not identified. In order to achieve such separation certain modifications to the completion process have been suggested \cite{clarke}.}  The $b$-boundary has also given some results that link the geometry of principal bundles with that of the base manifold \cite{stahl} and with non-commutative geometry \cite{keller}. Moreover, it has been shown that in four dimensions the Friedmann-Robertson-Walker and Schwarzschild $b$-completion $\partial {\mathcal M}$ is a point \cite{amores1, clarke, bosshard}.

The notion of $b$-incomplete spaces allows us to describe incomplete curves in manifolds with connections. Our initial motivation to study this, was to develop the language to describe pathologies in the geometry as we approach points that in some sense are ``boundary points" of the manifold. One can describe how the main manifestation of gravity in GR, the curvature of the manifold, can behave along $b$-incomplete curves. This is the scheme proposed by Ellis and Schmidt to classify singularities \cite{ellis, ellis2}.
In particular, they defined that  if $p\in\partial\mathcal{M}$ and there is some scalar constructed from the tensors $g_{ab}$, $R^{a}_{\;bcd}$ and $r$-covariant derivatives of $R^{a}_{\; bcd}$ that does not behave in a $C^{0}$ way, then $p$ is a $C^{r}$ \emph{scalar} singularity. Using this definition we have the following result

\begin{Theorem}
Let $\overline{\gamma}(s):[0,a]\rightarrow O(\mathcal{M}) $ be a lift from a curve $\gamma(t):[0,b]\rightarrow\mathcal{M}$ such that $\gamma(b)\in\partial\mathcal{M}$. If $\overline{R}\rightarrow -\infty$ as $s\rightarrow a$ then $|{R}|\rightarrow \infty$ as $t\rightarrow b$ and $\gamma(b)$is  a scalar singularity.
\end{Theorem}

The proof follows directly from Eq. \ref{ricci}.

Notice that the hypothesis of this theorem together with the hypothesis of any of the Hawking and Penrose theorems gives a singularity theorem in which it is guaranteed that curvature blow up exist. This is in contrast with the usual singularity theorem in which only geodesic incompleteness can be shown.

  The theorem above and the singularity theorems implicitly assume a characterisation of singularities in terms of  incomplete curves. This notion of singularity captures the idea that there are `obstructions' within the history of point-like observers.
In the future one would like extending those theorems to relate these obstructions  to curvature blow-up and ill-possessedness of initial value problems of field equations. This approach constitutes most of the research program on the Strong Cosmic Censorship 
conjecture \cite{Dafermos, Choquet}, the idea behind generalised hyperbolicity \cite{generalized, VW, ys2, ys} and field
regularity \cite{wald, kay, Marolf, IH, conical, ys1}.

  % Nevertheless, whenever the Schmidt metric has finite volume the geometric structure can be almost obtained through the fundamental group $\pi_{1}(\overline{O(\mathcal{M})})$ (see \cite{hatcher} for a precise definition of $\pi_{1}$ and \cite{afw} for the relationship between $\pi_{1}$ and the finite volume condition). For higher dimensions, the dimension of $\overline{O(\mathcal{M})}$ is now $n+\frac{n(n-1)}{2}>5$ and no classification is possible \cite{mav}. 

%-------------------------------------------------------------------------------------------------------------------------------------------------------
% ACKNOWLEDGEMENTS SECTION
%-------------------------------------------------------------------------------------------------------------------------------------------------------

\section*{Acknowledgments}
Y.S.S and C.M acknowledge funding support from CONACyT. The authors also thank  James Vickers,  Didier Solis and  Oscar Palmas for comments on previous drafts of the paper.

%-------------------------------------------------------------------------------------------------------------------------------------------------------
% APPENDIX SECTION
%-------------------------------------------------------------------------------------------------------------------------------------------------------

\appendix

%-----------------------------------------------------------------------------------------------
% THE RIEMANNIAN CASE
%-----------------------------------------------------------------------------------------------

\section{The Riemannian case}\label{s1}

As it was mentioned in Sec.\ \ref{1+1}, in the Riemannian case there are three conformally distinctly connected Riemann surfaces (the disc, the plane and the sphere). Moreover, in this case the fibres in $O({\mathcal{M}})$ are $SO(n,\mathbb{R})$ which is a compact group. Below we give the Schmidt metric for the general case of $1+1$ Riemannian manifolds and compute the curvature scalar for the disc, the sphere and the hyperbolic plane. 

In the Riemannian case it is a well know fact that if $\mathcal{M}$ is a 2-D manifold with a Riemannian metric we can find coordinates $(v,w)$ which transform locally the line element of the metric to a conformally flat form. Therefore, we have
\begin{equation}
  ds^{2}=\Omega^{2}(v,w)(dv^{2}+dw^{2}).
\end{equation} 

 An orthonormal basis is given by the vector fields
\begin{align}\label{basis}
 E_{1}=&\frac{1}{\Omega}\frac{\partial}{\partial{v}},\quad{\rm and} \\
  E_{2}=&\frac{1}{\Omega}\frac{\partial}{\partial{w}}. 
\end{align}
Any other orthonormal basis is constructed as a linear combination of \eqref{basis} as  
\begin{align}\label{basis2}
  \tilde{E}_{1}=&\cos\chi \frac{1}{\Omega}\frac{\partial}{\partial{v}}+\sin\chi \frac{1}{\Omega}\frac{\partial}{\partial{w}}\qquad {\rm and}\\
\tilde{E}_{2}=&\cos\chi \frac{1}{\Omega}\frac{\partial}{\partial{w}}+\sin\chi\frac{1}{\Omega}\frac{\partial}{\partial{v}}
\end{align}
for some $\chi\in\mathbb{R}$. Let us notice that the coefficients of basis \eqref{basis2} with respect to $\frac{\partial}{\partial v},\frac{\partial}{\partial w}$ define a unique matrix $\beta $ and its inverse $\beta ^{-1}$:
\begin{align}
\beta =\frac{1}{\Omega}
 \begin{pmatrix}
\cos\chi & -\sin\chi\\
\sin\chi & \cos\chi \\
   \end{pmatrix}
\quad{\rm and}\quad
 \beta^{-1}={\Omega}
   \begin{pmatrix}
\cos\chi & \sin\chi\\
-\sin\chi & \cos\chi \\
 \end{pmatrix}.
\end{align} 

Notice the main difference with the Lorentzian case in the definition of the matrix $\beta$.

The Schmidt metric $\overline{g}$ on $O({\mathcal{M}})$ is given by 
\begin{equation}\label{Schmidtmetric1}
\overline{g}(X,Y):\varpi(X)\cdot\varpi(Y)+\theta(X)\cdot\theta(Y),  
\end{equation}
for $X,Y\in TO(\mathcal{M})$.  
where   
\begin{align}
\theta(\dot{\gamma}) =  
\Omega \begin{pmatrix}
\dot{v}\cos\chi-\dot{w}\sin\chi\\
\dot{v}\sin\chi+\dot{w}\cos\chi \\
 \end{pmatrix}
\end{align} 
and 
\begin{align}{\varpi}(\dot{\gamma}) = 
 \begin{pmatrix}
0 & -\left(\dot{\chi}+\frac{1}{\Omega}((\partial_{v}{\Omega})\dot{w}+(\partial_{w}{\Omega})\dot{v})\right)\\
\dot{\chi}+\frac{1}{\Omega}((\partial_{v}{\Omega})\dot{w}+(\partial_{w}{\Omega})\dot{v})& 0 \\
 \end{pmatrix}
  \end{align}
giving the line element for the Schmidt metric:
\begin{align}\label{smri}
  ds^{2}=\Omega^{2}(v,w)(dv^{2}+dw^{2})+\left( d\chi+\frac{1}{\Omega}\left(\frac{\partial\Omega}{\partial v}dw+\frac{\partial\Omega}{\partial w}dv\right)\right)^{2}
\end{align}

\subsection*{The plane}
The euclidean metric on the plane is given by the line element
\begin{equation}
ds^{2}=dv^{2}+dw^{2} 
\end{equation} 
which is characterised by $R=0$. 

Then using Eq.\ \eqref{smri} we have that the line element for the corresponding Schmidt metric is
\begin{equation}
ds^{2}=dv^{2}+dw^{2}+d\chi^{2}  
\end{equation} 
which is just the flat metric in $O({\mathcal{M}})$ so we have $\overline{R}=0$ which violates the bound given by  Eq.\ref{ricci}.

\subsection*{The sphere}
The round metric on the sphere is given by the line element
\begin{equation}\label{rsphere}
ds^{2}=d\theta^{2}+\sin^{2}(\theta)d{\varphi}^{2}.
\end{equation} 
Eq.\ \eqref{rsphere} can be expressed in terms of isothermal coordinates $(v,w)$ as
\begin{equation}
ds^{2}=\frac{1}{\cosh^{2}v}(dv^{2}+dw^{2}). 
\end{equation} 
This metric is characterised by $R=1$. 

In a similar manner as we did for the plane metric we use Eq.\ \eqref{smri} to get the line element for the Schmidt metric:
\begin{equation}
ds^{2}= \frac{1}{\cosh^{2}v}(dv^{2}+dw^{2}) + (d \chi - \tanh(v) dw)^{2},
\end{equation} 
with curvature scalar $\overline{R}=3/2$. Notice that in this case the curvature scalar is positive which for Lorentzian manifolds can not happen as a result of Eq.\ref{ricci}.

\raggedright

%-------------------------------------------------------------------------------------------------------------------------------------------------------
% BIBLIOGRAPHY SECTION
%-------------------------------------------------------------------------------------------------------------------------------------------------------

%\bibliographystyle{apalike}
%\bibliography{biblio}

\begin{thebibliography}{1}

\bibitem{hawking1} S. W. Hawking and G. F. R. Ellis,
\textit{The Large Scale Structure of Space-Time.} 1st Edition,
Cambridge University Press, 1973.

\bibitem{senovilla} J. M. M. Senovilla.
Singularity Theorems and Their Consequences,
\textit{General Relativity and Gravitation.}
\textbf{30} (1998) 701--848.

\bibitem{sch} B. G. Schmidt.
A new definition of singular points in general relativity,
\textit{General Relativity and Gravitation.}
\textbf{1} (1971), 269--280.

\bibitem{confo} J. Frauendiener.
Conformal Infinity,
\textit{Living Reviews in Relativity.}
\textbf{7} (2004) 1.

\bibitem{flores2} J. L. Flores, J. Herrera and M. S{\'a}nchez.
The b-completion of the Friedmann space,
\textit{Advances in Theoretical and Mathematical Physics.}
\textbf{15} (2011), 991--1057.

\bibitem{scott} S.~M. Scott and P. Szekeres.
The abstract boundary---a new approach to singularities of manifolds,
\textit{Journal of Geometry and Physics.}
\textbf{13} (1994) 223--253.

\bibitem{scott1} B. E. Whale, M. J. S. L. Ashley and S. M. Scott.
Generalizations of the abstract boundary singularity theorem,
\textit{Classical and Quantum Gravity.}
\textbf{32} (2015), 135001.

\bibitem{low} R. J. Low,
\textit{''The Space of Null Geodesics (and a New Causal Boundary)?? in Analytical and Numerical Approaches to Mathematical Relativity}
Edited by J. Frauendiener, Domenico J.W. Giulini and V. Perlick,
Springer Berlin Heidelberg, 2006, pp.35--50.

\bibitem{ads} E. Witten.
Anti-de Sitter space and holography,
\textit{Advances in Theoretical and Mathematical Physics.}
\textbf{2} (1998) 253--291.

\bibitem{marolfgrg} D. Marolf and S. F. Ross.
A new recipe for causal completions,
\textit{Classical and Quantum Gravity.}
\textbf{20} (2003), 4085.

\bibitem{flores3} J. L. Flores, J. Herrera and M. S{\'a}nchez.
Gromov, Cauchy and causal boundaries for Riemannian, Finslerian and Lorentzian manifolds,
\textit{Memoirs of the American Mathematical Society.}
\textbf{226} (2013), No.1064.
\bibitem{differe} S. Kobayashi and K. Nomizu,
\textit{Foundations of Differential Geometry.}
Wiley, 1996.



\bibitem{gibbons} G.~W. Gibbons.
Quantized fields propagating in plane-wave spacetimes,
\textit{Communications in Mathematical Physics.}
\textbf{45} (1975) 191--202.

\bibitem{palmas} M. do Carmo,
\textit{Riemannian Geometry.}
Birkh{\"a}user, 1992.

\bibitem{kundt} W. Kundt.
Note on the completeness of spacetimes,
\textit{Zeitschrift fur Physik.}
\textbf{172} (1963) 488--489.

\bibitem{geroch} R. Geroch.
What is a singularity in general relativity?,
\textit{Annals of Physics.}
\textbf{48} (1968) 526--540.

\bibitem{amores2} A. M. Amores and M. Gutierrez.
Construction of examples of b-completion,
\textit{Nonlinear Analysis: Theory, Methods \& Applications.}
\textbf{47} (2001), 2959 -- 2970.

\bibitem{dodson} C.~T.~J. Dodson.
Space-Time Edge Geometry,
\textit{International Journal of Theoretical Physics.}
\textbf{17} (1978) 389--504.


\bibitem{beem} J. K. Beem.
Some examples of incomplete space-times,
\textit{General Relativity and Gravitation.}
\textbf{7} (1976) 501--509.

\bibitem{isothermal} T. Weinstein,
\textit{An Introduction to Lorentz Surfaces.}
De Gruyter, 1996.




\bibitem{bosshard} B. Bosshard.
On the b-boundary of the closed Friedman-model,
\textit{Communications in Mathematical Physics.}
\textbf{46} (1976) 263--268.

\bibitem{jon} R. A. Johnson.
The bundle boundary in some special cases,
\textit{Journal of Mathematical Physics.}
\textbf{18} (1977) 898--902.

\bibitem{clarke} C. T. J. Dodson.
Space-time edge geometry,
\textit{International Journal of Theoretical Physics.}
\textbf{17} (1978) 389--504.








\bibitem{Sta} F. St{\aa}hl.
Degeneracy of the b-Boundary in General Relativity
\textit{Communications in Mathematical Physics.}
\textbf{208} (1999), 331--353.

\bibitem{flores1} J. L. Flores, J. Herrera and M. S{\'a}nchez.
Hausdorff separability of the boundaries for spacetimes and sequential spaces,
\textit{Journal of Mathematical Physics.}
\textbf{57} (2016), 022503.


\bibitem{stahl} F. St{\aa}hl.
The Geometry of the Frame Bundle over Spacetime,
\textit{ArXiv:0006049.}
(2000).

\bibitem{keller} M. Heller, Z. Odrzygozdz, L. Pysiak and W. Sasin.
Anatomy of malicious singularities,
\textit{Journal of Mathematical Physics.}
\textbf{48} (2007), 092504.

\bibitem{amores1} A. M. Amores and M. Gutierrez.
The b-completion of the Friedmann space,
\textit{Journal of Geometry Physics.}
\textbf{29} (1999), 177--197.




\bibitem{ellis} G.~F.~R. Ellis and B.~G. Schmidt.
Classification of singular space-times,
\textit{General Relativity and Gravitation.}
\textbf{10} (1979) 989--997.

\bibitem{ellis2} G.~F.~R. Ellis and B.~G. Schmidt.
Singular space-times,
\textit{General Relativity and Gravitation.}
\textbf{8} (1977) 915--953.



\bibitem{Dafermos} M. Dafermos.
The interior of charged black holes and the problem of uniqueness in general relativity,
\textit{Communications on Pure and Applied Mathematics.}
\textbf{58} (2005) 445--504.

\bibitem{Choquet} Y. Choquet-Bruhat,
\textit{General Relativity and the Einstein Equations.} 1st Edition,
OUP Oxford, 2009.

\bibitem{generalized} C. J. S. Clarke.
Generalized hyperbolicity in singular spacetimes,
\textit{Classical and Quantum Gravity.}
\textbf{15} (1998) 975.

\bibitem{VW} J.~A. Vickers and J.~P. Wilson.
Generalised hyperbolicity: hypersurface singularities,
\textit{ArXiv:0101018}.

\bibitem{ys2} Y. Sanchez Sanchez and J. A. Vickers.
Generalised hyperbolicity in spacetimes with Lipschitz regularity,
\textit{Journal of Mathematical Physics.}
\textbf{58} (2017) 022502.

\bibitem{ys} Y. Sanchez Sanchez and J. A. Vickers.
Generalised hyperbolicity in spacetimes with string-like singularities,
\textit{Classical and Quantum Gravity}
\textbf{33} (2016) 205002.

\bibitem{wald} R.~M. Wald.
Dynamics in nonglobally hyperbolic, static space-times,
\textit{Journal of Mathematical Physics.}
\textbf{21} (1980) 2802--2805.

\bibitem{kay} B.~S. Kay and U.~M. Studer.
Boundary conditions for quantum mechanics on cones and fields around cosmic strings,
\textit{Communications in Mathematical Physics.}
\textbf{139} (1991) 103--139.

\bibitem{Marolf} G.~T. Horowitz and D. Marolf.
Quantum probes of spacetime singularities,
\textit{Physical Review D.}
\textbf{52} (1995) 5670--5675.

\bibitem{IH} A. Ishibashi and A. Hosoya.
Who's afraid of naked singularities? Probing timelike singularities with finite energy waves,
\textit{Physical Review D.}
\textbf{60} (1999) 104028.

\bibitem{conical} J.~P. Wilson.
Generalized hyperbolicity in spacetimes with conical singularities,
\textit{Classical and Quantum Gravity.}
\textbf{17} (2000) 3199--3209.

\bibitem{ys1} Y. Sanchez Sanchez.
Regularity of curve integrable spacetimes,
\textit{General Relativity and Gravitation.}
\textbf{47} (2015) 80.




\end{thebibliography}

%--------------------------------------------------------------- END DOCUMENT --------------------------------------------------------------
\end{document}